# University of Oxford

**Petar Radanliev, BA Hons., MSc., Ph.D.**
POSTDOCTORAL RESEARCH ASSOCIATE

DEPARTMENT OF
ENGINEERING
SCIENCE

UNIVERSITY OF
OXFORD

# The Rise and Fall of Cryptocurrencies: Defining the Economic and Social Values of Blockchain Technologies, assessing the Opportunities, and defining the Financial and Cybersecurity Risks of the Metaverse.

Dr Petar Radanliev

Department of Engineering Science, University of Oxford, United Kingdom:
petar.radanliev@eng.ox.ac.uk

**Abstract:**

> **This paper contextualises the common queries of "why is crypto crashing?" and "why is crypto down?", the research transcends beyond the frequent market fluctuations to unravel how cryptocurrencies fundamentally work and the step-by-step process on how to create a cryptocurrency.**

The study examines blockchain technologies and their pivotal role in the evolving Metaverse, shedding light on topics such as how to invest in cryptocurrency, the mechanics behind crypto mining, and strategies to effectively buy and trade cryptocurrencies. Through an interdisciplinary approach, the research transitions from the fundamental principles of fintech investment strategies to the overarching implications of blockchain within the Metaverse. Alongside exploring machine learning potentials in financial sectors and risk assessment methodologies, the study critically assesses whether developed or developing nations are poised to reap greater benefits from these technologies. Moreover, it probes into both enduring and dubious crypto projects, drawing a distinct line between genuine blockchain applications and Ponzi-like schemes. The conclusion resolutely affirms the continuing dominance of blockchain technologies, underlined by a profound exploration of their intrinsic value and a reflective commentary by the author on the potential risks confronting individual investors.

# University of Oxford

**Petar Radanliev, BA Hons., MSc., Ph.D.**
POSTDOCTORAL RESEARCH ASSOCIATE

**Note about the author:**

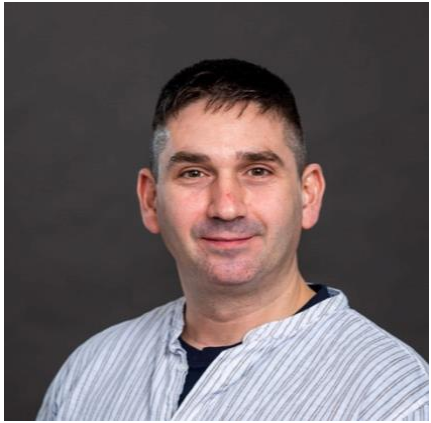

Dr Petar Radanliev began his career as a penetration tester for the military and the defence industry. Subsequently, he transitioned to cyber risk management in the finance sector. After a decade in defence and finance, he re-joined the academic realm, obtaining his Ph.D., MSc, and BA (Hons) from the University of Wales. Before joining Oxford, Petar undertook postdoctoral research projects at Imperial College London, the University of Cambridge, MIT, and the University of North Carolina. He held positions as a Prince of Wales Innovation Scholar at the University of Wales and as a Fulbright Scholar at both MIT and the University of North Carolina. His research areas encompass Artificial Intelligence, Generative Pre-trained Transformers (GPT), Cybersecurity, Blockchain Technologies, Cryptocurrencies, and the Internet of Things. Recent research specialisations include the Software Bill of Materials (SBOM) and the Vulnerability Exploitability Exchange (VEX).

**Funding**: This work has been supported by the PETRAS National Centre of Excellence for IoT Systems Cybersecurity, which has been funded by the UK EPSRC [under grant number EP/S035362/1]; the Software Sustainability Institute [grant number: EP/S021779/1]; and by the Cisco Research Centre [grant number CG1525381].

**Acknowledgements**: Eternal gratitude to the Fulbright Commission.

# University of Oxford

**Petar Radanliev, BA Hons., MSc., Ph.D.**
POSTDOCTORAL RESEARCH ASSOCIATE

## 1. Introduction.

In recent years, cryptocurrency has emerged as a significant and often contentious component of the financial landscape. This article delves deep into the complex world of digital currencies, clarifying the methods of investing in these volatile assets and the intricate mechanisms of crypto mining vs crypto staking. We explore the various platforms and methodologies for purchasing cryptocurrency, whilst also casting insights on the frequent price fluctuations observed in the market, analysing both the technical and external factors leading to periodic crashes and downturns. To provide a comprehensive understanding, the underlying technology that powers these digital tokens is broken down, offering insights into the decentralised nature of blockchain-based assets. This study navigates the steps and challenges involved in creating a new cryptocurrency and guide readers through the intricacies of trading and maximising potential returns in the crypto market.

This article provides an up-to-date assessment of the current state of cryptocurrencies, examining both their values and associated risks within the realm of blockchain technology. With the exponential growth of over 20,000 crypto projects, the study offers a snapshot of the landscape in 2023, tracing the historical trajectory from Satoshi's ground-breaking paper on decentralised blockchains to the present day. The study clarifies the distinctions between cryptocurrencies and blockchain technologies while exploring pertinent research questions: Is blockchain technology an innovation or an obsolete concept? What significant risks do cryptocurrencies entail? Are the potential societal and economic benefits worth pursuing?

In light of the immense growth observed in the crypto market, encompassing over 20,000 projects, this investigation offers a timely snapshot of the landscape in 2023 while providing a historical account from the seminal work of Satoshi on decentralised blockchains to the present day. The study objectives extend beyond merely examining values and risks associated with cryptocurrencies; we also aim to delineate the nuances distinguishing cryptocurrencies from blockchain technologies. Guided by a series of research questions, we seek to ascertain the nature of blockchain technology as either an innovation or an outdated concept, identify significant risks posed by cryptocurrencies, evaluate their potential value to society and the economy, and determine which countries are likely to benefit the most from these technologies. Additionally, we endeavour to shed light on the survivability prospects of various blockchain projects, thereby addressing whether the allure surrounding them is hype or a substantial opportunity. While acknowledging the presence of numerous fraudulent crypto ventures, often functioning as Ponzi schemes, our focus primarily lies in exploring the genuine use cases of blockchain projects. Ultimately, this research study culminates in a resolute affirmation that blockchain technologies are here to stay, substantiated by an extensive discussion of their intrinsic value. Furthermore, we re-evaluate vital risks, including a personal statement from the author regarding the perils faced by individual investors.

### 1.1. Historical context

This article has been around 13 years in the making. I started writing this article back in 2009, with the emergence of Bitcoin [1]. The report is influenced by points of view that existed before 2009 and have long been forgotten, such as the fear of Bitcoin

# University of Oxford

**Petar Radanliev, BA Hons., MSc., Ph.D.**
POSTDOCTORAL RESEARCH ASSOCIATE

(BTC) legality, the ethics of decentralised control, and the 'ethical impact of cryptocurrencies as morally beneficial, detrimental, and ambiguous' and the 'antisocial use for shadow banking and transactions in the 'dark net' and cryptocurrencies' effect on inflation and deflation' [2]. Although some of these viewpoints have subdued, and cryptocurrencies are legal to own and operate in many countries, many of these fears remain among early adopters. The article discusses the values and risks of a few selected Blockchain projects based on real-world value and their potential to contribute to the future of society.

The main research areas reviewed in this study are included in Figure 1, which also helps visualise the main areas of interest in the year 2023 in the Metaverse concept.

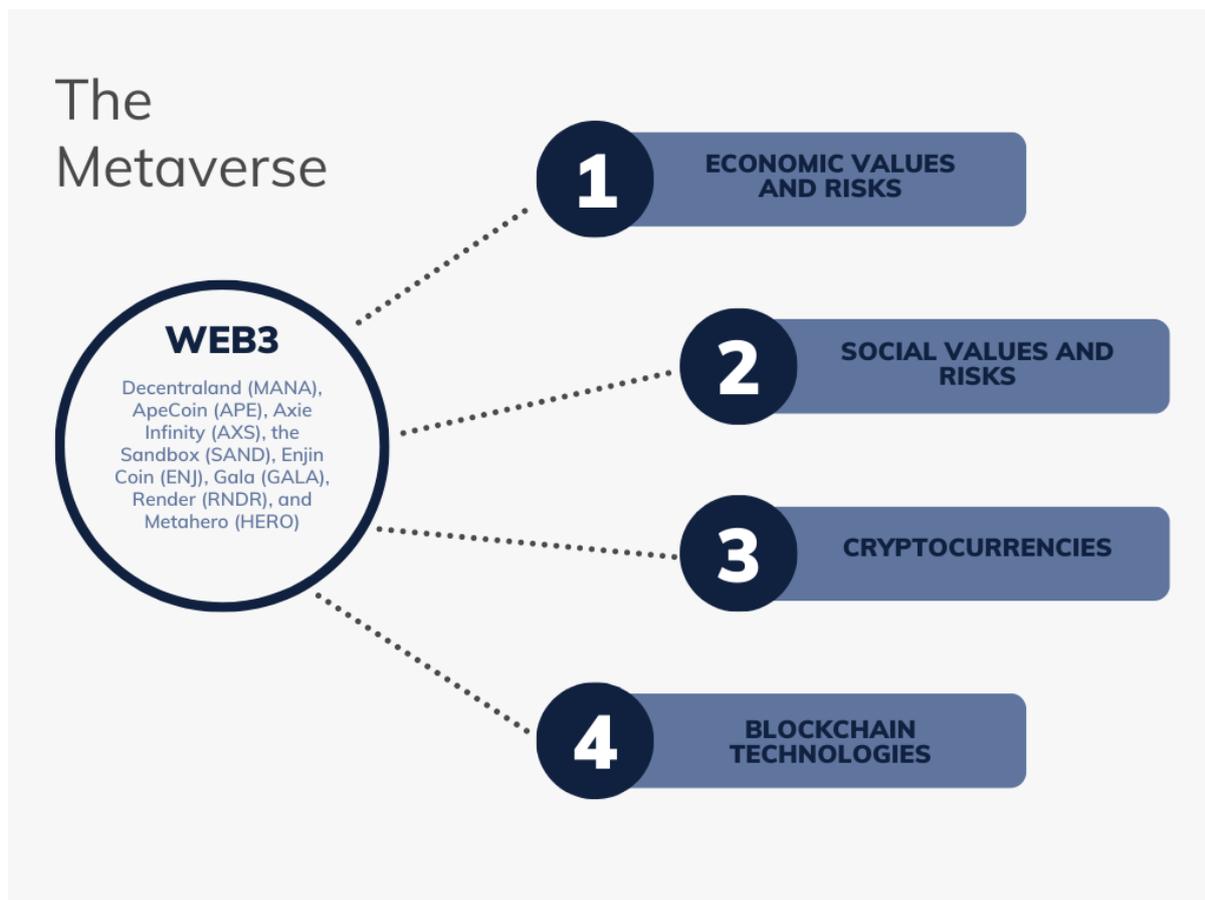

*Figure 1: Overview of the research topics discussed.*

## 1.2. Historical overview of Blockchain technologies

The first whitepaper on Bitcoin emerged at the peak of the financial crisis in 2008 [1] and it promoted the idea of a different economic system that is not dependent on a trusted third party. In 2009, the main concern was the legality of such currencies and projects. Bitcoin was considered a mechanism for criminals and drug dealers to bypass the legal banking system. Between 2020 and 2022, investment in cryptocurrencies (crypto) exploded. Despite the collapse of Terra Luna, the FTX exchange in 2022, and the overall downturn in the market for crypto, crypto investments have continued.

Cryptocurrencies, or digital assets, serve as virtual currencies underpinned by cryptography, offering heightened security and privacy measures. They often operate decentralised, distinguishing them from traditional forms of money. Notably, a defining characteristic of cryptocurrencies is their purported immunity to governmental or institutional control. However, the validity of this claim is subject to debate, as the potential for a single government, such as the United States, or a dominant institution, like BlackRock, to acquire a controlling stake in a cryptocurrency undermines this principle. It is crucial to recognise that while Bitcoin remains the most widely recognized cryptocurrency, numerous other cryptocurrencies exist in the market. Cryptocurrencies can facilitate the purchase of goods and services and are actively traded on various online platforms. It is worth noting, however, that none of these platforms are subject to comprehensive regulation. This regulatory vacuum presents a unique opportunity for countries like the UK, which seek to establish their presence internationally, particularly in light of recent challenges stemming from the impacts of Covid-19 and Brexit.

One argument for pursuing Cryptocurrencies as a solution to financial risk is '*that the main indicators to improve financial development should enhance the process of bank lending and equity market development'* [3] Second argument is that smart contracts and metaheuristic can help with securing the quality-of-service and even help with the '*cost-efficient scheduling of medical-data processing'* [4], which seems of crucial importance for the medical systems in the UK – after the Covid shock to the National Health Service (NHS) [5]–[7].

In summary, pursuing cryptocurrencies as a solution to financial risk is supported by the potential to improve economic development indicators, such as bank lending and equity market development, while simultaneously addressing crucial challenges in the healthcare sector. By harnessing the capabilities of smart contracts and metaheuristic techniques, cryptocurrencies offer opportunities for enhancing the quality-of-service and cost-efficiency of medical data processing, which is of utmost importance for the UK's healthcare systems in the aftermath of the Covid-19 pandemic. These advancements can potentially bring about transformative changes, promoting financial stability and bolstering the resilience of the healthcare sector.

## 1.3. Brief History of Cryptocurrencies.

Cryptocurrencies (crypto) are digital assets, or more precisely, a set of digital currencies that emerged with the release of Bitcoin in 2009[1]. As of January 2021, there are over 4,000 in circulation, some cryptos with minimal trading volume (or not at all). Cryptocurrencies are traded as digital 'tokens' or 'coins' on a distributed and decentralised ledger. Bitcoin leads on market capitalisation, but other cryptos are trying to break into the market by providing different and improved services to Bitcoin. Some of the other cryptos - also known as 'altcoins', i.e., all cryptocurrencies other than Bitcoin – are used to create a decentralised financial system, e.g.,

---

[1] https://bitcoin.org/en/

# University of Oxford

**Petar Radanliev, BA Hons., MSc., Ph.D.**
POSTDOCTORAL RESEARCH ASSOCIATE

Ethereum[2], with the ability to handle more transactions, e.g., Dogecoin[3], or just use different consensus algorithms, e.g., Cardano[4]. Cryptocurrencies remain highly volatile, and without central control, a single statement by Elon Musk has been sufficient to trigger a spike in search trends (as seen in Figure 2) and attract significant interest in news media.

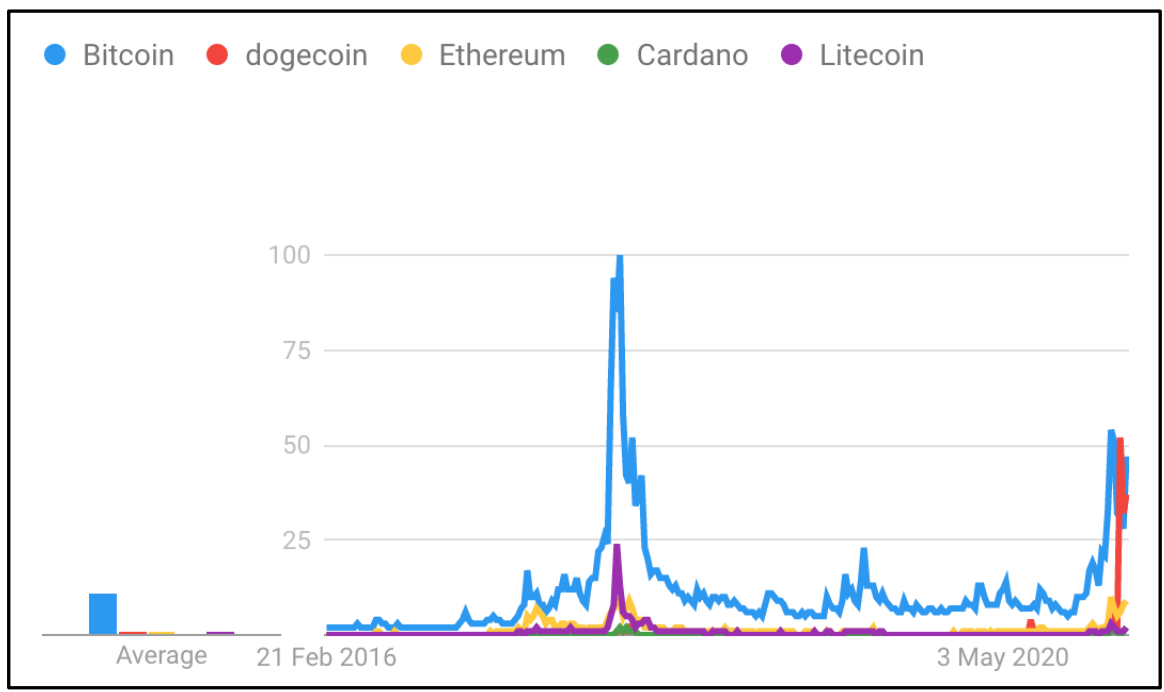

*Figure 2: Comparison of search trends on some of the most popular cryptos (data from 2020 – before the last prominent bull run in crypto)*

Although the search trends can spike about specific cryptos, i.e., Bitcoin and Dogecoin in Figure 2, a dynamic equicorrelation [8] shows a contagious correlation (a mutual relationship or connection) effect between cryptocurrencies (i.e., when Bitcoin crashes, altcoins follow), and the exact relationship is also connected between Bitcoin and NASDAQ. This disincentivises diversification into multiple cryptocurrencies, but a more robust analysis with a value-at-risk [9]–[13] and expected shortfall must be computed to confirm this. Looking at the spikes in search trends after the announcement from Elon Musk on Bitcoin purchase and comparing them to the price cap of Bitcoin and Dogecoin, it almost resembles a specific behaviour in crypto markets.

The user behaviour in some crypto (e.g., Ethereum) appears more stable. In contrast, the behaviour of users in Bitcoin seems more speculative, with fluctuations based on market trends, followed by more considerable sell-outs when the market is down [14]. This is not to say that influential people cannot manipulate the stock

---

market, but the point here is that stock market investors are protected by regulations which don't yet exist in crypto markets.

Figure 3 shows that 'interest over time' has generally changed for Bitcoin and Cryptocurrencies – including Altcoins. While the interest in Google Trends for Bitcoin peaked in the 2018 bull run, the interest in cryptocurrencies, as a search trend, has increased (to a new 100%) in the 2021 bull run.

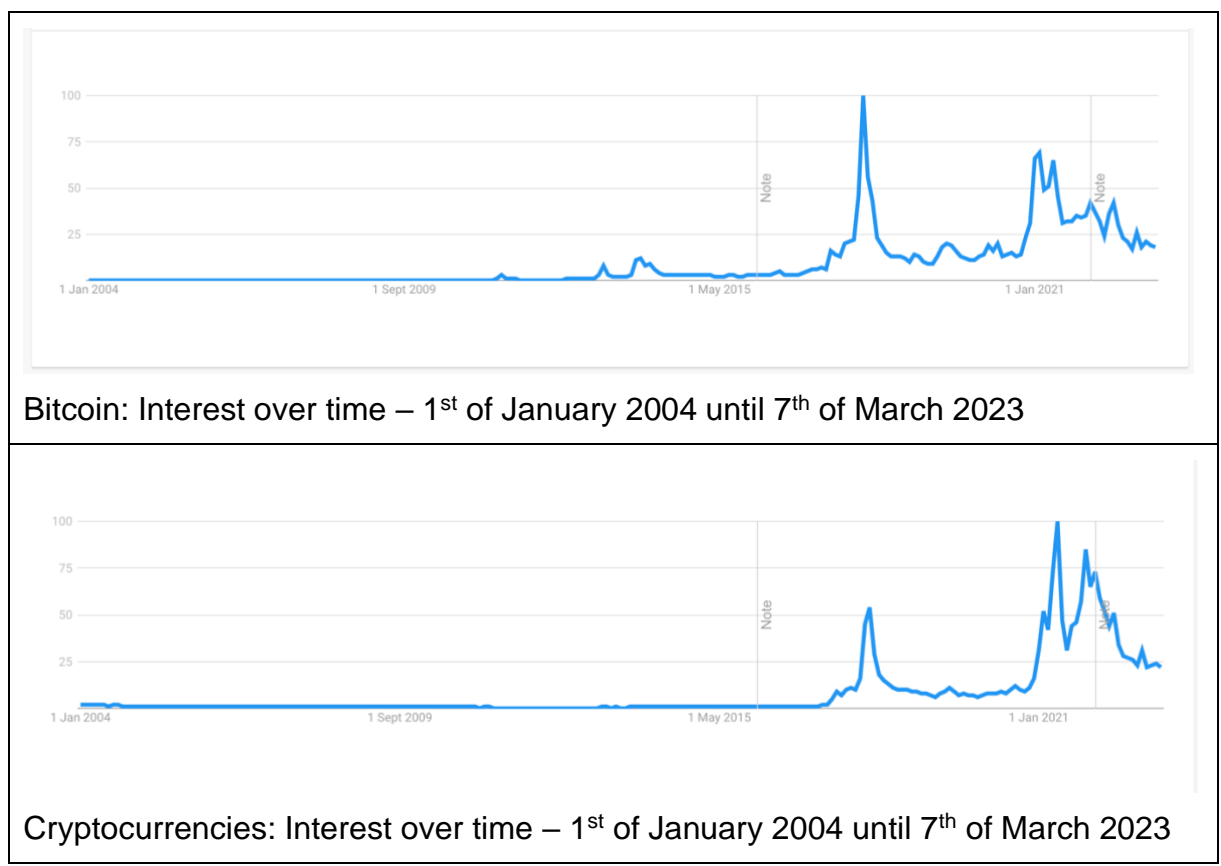

Bitcoin: Interest over time – 1st of January 2004 until 7th of March 2023

Cryptocurrencies: Interest over time – 1st of January 2004 until 7th of March 2023

*Figure 3: Interest over time: Bitcoin vs Cryptocurrencies (including Altcoins)*

Crypto users can be categorised according to their actions and resources, e.g., fortune, hunter, and idealist [15]. Still, the bigger question is, can we classify cryptocurrencies into taxonomies and forecast individual Altcoin categories' future success or failure? Another important topic discussed in this article is related to cryptocurrencies, the idea of Central Bank Digital Currencies (CBDCs) [16]–[18], and Friedrich von Hayek's theory of private money, which some experts argue could lead to national currencies being replaced by *'the currency of the digital platform'* [19]. Another point on centralisation is that even cryptocurrencies that are meant to be very decentralised, such as Ethereum (ETH), and stablecoins, such as UDSC, can be (in reality) very centralised and potentially controllable.

### 1.4. Research questions and structure.

The research aims, objectives, questions, and goals motivating this article can be defined in one sentence as the current and future values and risks associated with blockchain projects. The areas of focus encouraging this study include:

1. To assess the potential of blockchain technology as a catalyst for innovation in future iterations of the internet and Web3 and to determine whether it represents a cutting-edge technology or an outdated concept in the context of financial invention.
2. To comprehensively analyse and evaluate the risks associated with cryptocurrencies, including but not limited to market volatility, security vulnerabilities, regulatory challenges, and potential illicit activities, to understand the risk landscape within financial innovation better.
3. Investigating and identifying the intrinsic values derived from cryptocurrencies, considering their potential impact on financial systems, economic growth, financial inclusion, transaction efficiency, and transparency, thus contributing to understanding their value proposition within financial innovation.
4. Conduct a thorough analysis of different blockchain projects and their underlying technologies to identify critical factors that contribute to their long-term viability and survival in the dynamic landscape of financial innovation, thereby providing insights into the sustainability of blockchain projects.
5. To explore and identify blockchain projects that have the potential to further the development and advancement of already developed countries' financial systems, addressing specific areas such as efficiency, security, transparency, financial inclusion, and regulatory compliance, thus providing valuable insights into the role of blockchain in enhancing financial innovation in advanced economies.
6. To examine and identify blockchain projects that can foster development and progress in developing countries, taking into consideration their unique challenges and needs, including financial inclusion, access to capital, remittances, land registries, supply chain management, and government services, thereby contributing to the understanding of how blockchain can drive financial innovation in developing economies.
7. To critically evaluate the feasibility and implications of regulating cryptocurrencies and blockchain projects, including both national and international regulatory frameworks, to assess the potential benefits, challenges, and trade-offs associated with effective regulation in the context of financial innovation. This research objective seeks to contribute to the ongoing discourse on appropriate regulatory approaches for cryptocurrencies and blockchain technology.

The article follows a traditional (standard) structure, starting with chapter one (1) Introduction, chapter two (2) Methodology, then (3) the research engages with a Case study review of secondary data sources, and compliments the case study with a new chapter (4) that comprises a survey review. The article ends with a chapter (5) discussion and (6) Conclusion.

## 1.5. Crypto regulations.

Given that the US has already started crafting new regulations on cryptocurrencies [16] and after the markets collapsed in 2022, the European Union also started catching up with rules [20] the UK recently started initial consultation plans to regulate cryptocurrencies [20]. It is difficult to determine if the proposed methods are positive or negative. Still, the consultation generally builds upon previous HM

Treasury proposals on stablecoins. If the plan is to develop a UK stablecoin (USD and GBR), the talk could prove positive for ensuring stability in cryptocurrency markets. However, the proposed regulation suggests that cryptocurrencies should be overseen by the Financial Conduct Authority (FCA), and not all crypto is just a simple cryptocurrency. Ethereum (for example) is a virtual computer and can perform far more tasks than just transact payments. The consultation also suggests that all UK-based cryptocurrency firms should have anti-money laundering and KYC processes. This can be managed in the UK, but the regulation ignores that cryptocurrencies are designed to bypass such rules. The effectiveness of this regulation remains questionable, but it could help stabilise and legalise the trade of cryptocurrencies in the UK. In other words, these regulations could apply to large companies compliant with many different rules. Still, it's difficult to see how such laws would prevent money laundering on non-UK-based cryptocurrency firms.

## 1.6. Structure and novelty of the research study

This study offers a multifaceted exploration of blockchain technologies, investigating their economic, social, and wider implications within the Metaverse. The research highlights key themes from blockchain's impact across sectors to the fintech revolution. Contrasting with existing literature, this work broadens the discussion from the mechanics of cryptocurrency trading to the wider ramifications of blockchain, emphasising its financial and cybersecurity risks. The novelty of the research is further underscored by its interdisciplinary approach, merging the fundamental principles of fintech investment strategies, blockchain, and the metaverse. Such a comprehensive perspective positions this study uniquely, complementing and enriching the current academic discourse on cryptocurrency trading, machine learning potentials in financial sectors, and financial risk assessment methodologies. The research not only provides insights into the evolving digital landscape but also sets a new benchmark for future investigations at the crossroads of finance, technology, and societal constructs. Then, the study investigates whether developed or developing countries stand to gain more from these technologies and which blockchain projects are likely to endure in the long term. This review acknowledges the prevalence of dubious crypto projects, delving into the realm of Ponzi schemes yet ultimately shifting the focus towards the practical applications of blockchain projects. The study concludes with a resolute assertion that blockchain technologies are here to stay, supported by a comprehensive discussion of their inherent value. Moreover, the article reassesses critical risks, including a personal reflection from the author on the potential risks' individual investors face.

# 2. Methodology.

In this article, we present a robust research methodology, employing a combination of the case study method, survey review, and literature review to delve into the dynamic realm of financial innovation and its social implications. Drawing inspiration from established and time-tested guidelines such as Eisenhardt's 'Building Theories from Case Study Research' [21], Yin's recommendations on case study research, design, and methods [23], and the principles of Grounded theory [22], and the Grounded theory [23]–[25] Figure 4, provides a visual representation of the structured methodology, delineating areas of interest and focus. Our research methodology is thoughtfully divided into a dual emphasis on economic and social values. Recognising that assessing values, risks, and impact extends beyond purely economic metrics, we aim to elucidate the interplay between financial and social considerations.

A prominent concern driving the exploration of Web3 arises from the challenges posed by Web2, particularly regarding diminishing personal privacy. Web2 sceptics advocate for a transition to Web3 not solely for its decentralised financial framework but also for its potential to enhance individual privacy. Embracing Blockchain technologies in the Web3 design, along with the integration of virtual and augmented (mixed) reality in education, fosters opportunities for improved personal privacy, decentralised social media, and global community building of like-minded individuals.

By adopting this methodological approach, we aspire to offer a coherent and understandable analysis of the intricate landscape of financial innovation, elucidating the role of Web3 in addressing critical issues and shaping the future of finance, privacy, and social interactions. Through a multidimensional lens, we aim to contribute valuable insights into the potential benefits and challenges of embracing Web3, shedding light on its transformative impact on the digital landscape.

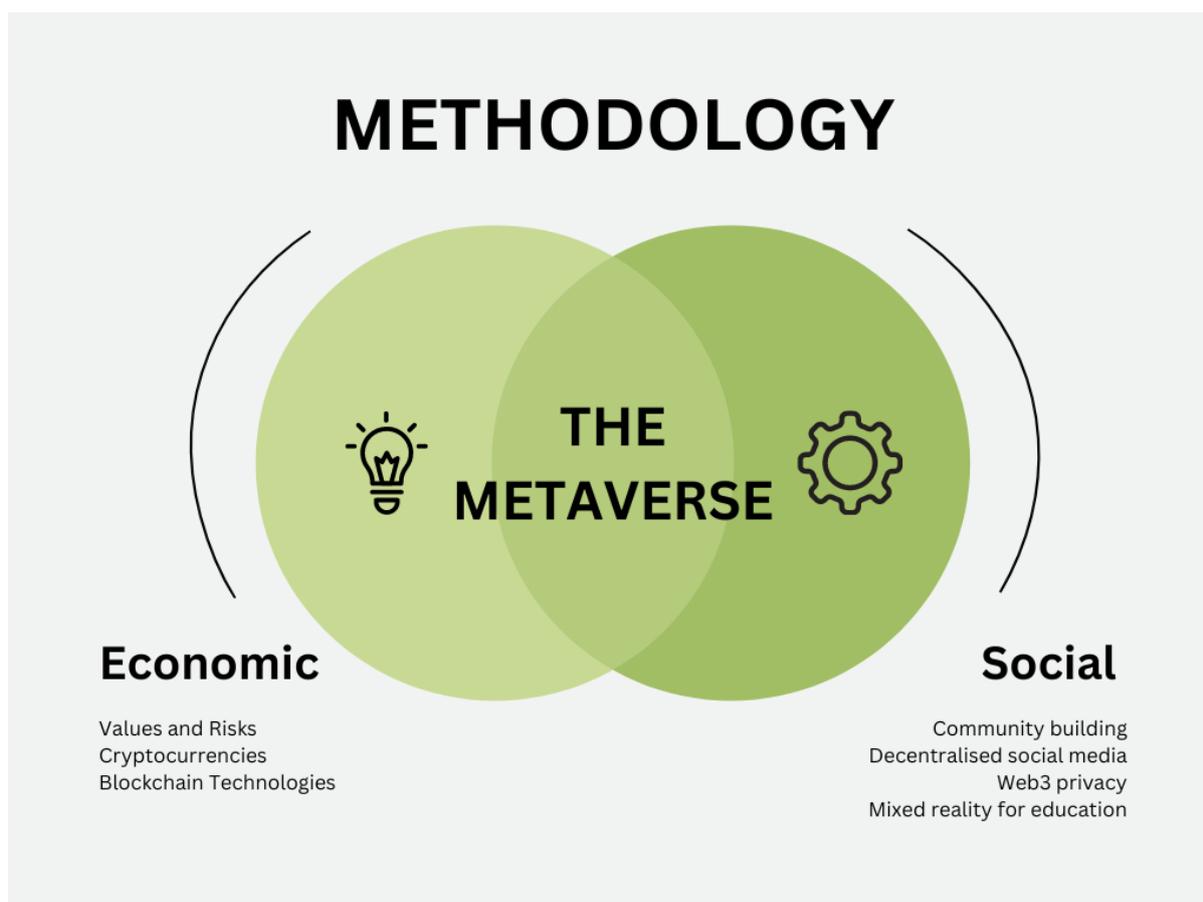

*Figure 4: The case study method and the study objectives*

Since the initial search on the Web of Science and Scopus didn't result in many records on the topics discussed in Figure 4, the search was expanded into a broader review of Google Scholar, IEEE, and other libraries. Although the initial stages focused on new technologies, the study expanded into the literature on ethics, trading, gambling, decentralised finance, and the Metaverse after the initial search.

## 2.1. Data sources.

The primary data collection included 20 case study interviews and three workshops. The secondary data collection included new and emerging forms of data (NEFD) that present new records of value in IoT-based intelligent contracts, including open data – (e.g., Open Data Institute); spatiotemporal data; high-dimensional data; time-stamped data; real-time data and big data. To analyse NEFD, the data collection strategy in this article applied *stratified sampling* and *random sampling* for comparative analysis of collected NEFD.

## 2.2. Contextualising this study with the existing body of literature

This section is focused on contextualisation of the findings and the novelty of the work in this study. This section compares the differences and similarities, including the findings and novelty, of this work, with eight articles closely related to the research area investigated in this study.

This study started with a systematic review of existing studies that used data records obtained from the Web of Science Core Collection [26], focusing on the broad landscape of blockchain's impact across sectors, highlighting key themes from economic benefits to the fintech revolution. Within this body of literature, this study on "The Rise and Fall of Cryptocurrencies" offers a complex and interesting exploration, delving deep into the intricacies of cryptocurrency and the broader implications in the metaverse. This article not only complements the identified themes but goes a step further by elucidating the financial and cybersecurity risks associated with blockchain technologies. This in-depth focus on the metaverse and its associated challenges underlines the novelty of this work, positioning it as a pivotal piece that bridges gaps in the current literature and offers fresh insights into the evolving digital realm.

This study delves into the intricacies of blockchain technologies, highlighting their economic and social values while scrutinising opportunities and risks inherent to the metaverse. In juxtaposition to the comprehensive survey on cryptocurrency trading by Fan Fang et al., [27] this work holds significance. While Fang and colleagues extensively map out the landscape of cryptocurrency trading, from sentiment-based research to the public nature of blockchain technology, our study on exploration underscores broader implications, particularly in relation to the metaverse's opportunities and challenges. This positions this new article uniquely, as it extends the dialogue beyond the mechanics of trading to the ramifications of blockchain technologies in shaping our digital futures. This new investigation into the metaverse, an area not deeply touched upon by Fang et al., accentuates the novelty and contemporaneity of this research, presenting a holistic view that complements and augments existing literature.

In the broader discourse surrounding the implications of cryptocurrencies, this study delineates the multi-dimensional societal and economic repercussions of blockchain innovations, setting an expansive stage to understand their systemic influences. By contrast, the research on "Forecasting and trading cryptocurrencies with machine learning under changing market conditions" [28] offers a deep dive into the predictive and profit-centric paradigms of these digital assets, specifically focusing on machine learning's applicability in this domain. The convergence of these studies is particularly enlightening. While the former casts a wide net, encompassing the multifaceted socio-economic interplays and the emergent realm of the metaverse, the latter drills down, examining the nuances of cryptocurrency trading amidst volatile market conditions. The overarching narrative of "The Rise and Fall of Cryptocurrencies," when viewed in tandem with the in-depth findings on ML's potential and challenges in deciphering cryptocurrency markets, offers a synergistic understanding. It enriches the academic conversation, providing a holistic and multi-layered exploration of the cryptocurrency ecosystem, from its foundational technologies to the Metaversal futures they portend.

In the contemporary milieu of group decision-making processes, the study titled "Soft consensus cost models for group decision-making and economic interpretations" [29]provides a robust foundation for understanding the nuances and costs of consensus-building among experts. This work unravels the complexity of striking a balance between achieving consensus and incurring associated costs with the introduction of the novel concept of soft minimum cost consensus. Their findings not only advance previous models but also delve into the economic underpinnings of such consensus mechanisms. Within this context, this study on "The Rise and Fall of Cryptocurrencies" emerges as a novel contribution to the discourse. This investigation transcends the traditional realms of consensus-building in decision-making contexts and dives deep into the multifaceted world of cryptocurrencies and the burgeoning metaverse. It not only seeks to define the intrinsic economic and social values underpinning blockchain technologies but also aspires to critically assess the opportunities and challenges presented by the metaverse. The novelty of this work lies in its ability to contextualise and intertwine the foundational principles of consensus models, as observed in the prior study, with the intricate dynamics of blockchain technologies. By comparing the core tenets of group decision-making and the rapidly evolving world of digital assets, this study offers fresh insights and sets a new precedent for interdisciplinary research in this domain.

In the dynamic world of financial technology, the pivotal research "Fintech investments in European banks: a hybrid IT2 fuzzy multidimensional decision-making approach" [30] pedantically evaluates the essence of Fintech investments in European banking services, elucidating key criteria and offering indispensable insights into optimal investment alternatives. Notably, the emphasis on payment and money-transferring systems as paramount Fintech investment alternatives presents a foundational benchmark in the industry. Within this rich tapestry of Fintech investigations, "The Rise and Fall of Cryptocurrencies: Defining the Economic and Social Values of Blockchain Technologies and Assessing the Opportunities, and Risks of the Metaverse" offers an innovative and integrative exploration of blockchain technologies and their interplay with the metaverse. By seamlessly weaving the threads of cryptocurrencies' economic and social values, this research transcends the conventional boundaries of financial technology discussions and ventures into the evolving realms of the Metaverse. The novelty of this study is underscored by its

# University of Oxford

**Petar Radanliev, BA Hons., MSc., Ph.D.**
POSTDOCTORAL RESEARCH ASSOCIATE

multidisciplinary approach, integrating the core tenets of Fintech investment strategies with the multifaceted nuances of blockchain and the metaverse. In doing so, it not only complements existing works but also carves a distinct niche in contemporary financial and technological scholarship, presenting fresh vistas and research trajectories in the interplay of Fintech, blockchain, and virtual realms.

In comparison with the research on the "Integrated Cluster Detection, Optimization, and Interpretation Approach for Financial Data," [31] which concentrates on the identification and interpretation of clusters within financial data to enhance decision-making processes, this study on cryptocurrencies offers a broader perspective. Where the former employs sophisticated methodologies to discern patterns in vast financial datasets, emphasising adaptivity, speed, and interpretability, our exploration into the realm of cryptocurrencies and the metaverse endeavours to contextualise these innovations within the larger socio-economic canvas. The novelty of our investigation lies in its interdisciplinary approach, bridging the gaps between finance, technology, and social constructs and presenting a holistic understanding of the current digital transformation. Thus, while the study offers invaluable tools for immediate financial data interpretation, our research equips readers with a comprehensive understanding of the evolving landscape of blockchain technologies and the metaverse, forecasting potential trajectories and their implications on our collective future.

This study casts a discerning eye on the emergent dynamics of blockchain technologies, pivoting on their socio-economic implications and the opportunities and risks they introduce in the digital realm of the metaverse. When compared with the study titled "Machine Learning Methods for Systemic Risk Analysis in Financial Sectors," [32] which delves into the potent intersections of machine learning and systemic risk within the financial sphere, our research offers a more expansive vista, encompassing not just the financial but also the societal reverberations of digital currencies and virtual spaces. While the referenced article meticulously maps the deployment of machine learning techniques in gauging financial systemic risk, underscoring the power of big data, network, and sentiment analyses in this quest, our study pivots on the broader implications and trajectory of blockchain and the metaverse. The novelty of our research resides in its comprehensive purview, encapsulating not just financial but also socio-cultural and technological dimensions. While the surveyed work offers invaluable insights into the detection and modulation of systemic risks through machine learning, our article contextualises cryptocurrencies and the metaverse within the evolving tapestry of our digital age, seeking to foretell and shape the contours of this transformation.

For a final contextualisation of the findings and novelty of the work in this study, we need to emphasise that in this study, the multifaceted impacts, and prospects of blockchain technologies, with a particular focus on cryptocurrencies, are scrutinised within both socio-economic dimensions and the expanding horizon of the metaverse. Contextualising this work against the study titled "Evaluation of clustering algorithms for financial risk analysis using MCDM methods," [33] which meticulously dissects the application of multiple criteria decision-making methods in ranking clustering algorithms for financial risk analyses reveals notable distinctions. While the latter delves deep into the computational methodologies employed to enhance precision in financial risk evaluation, our investigation spans a broader spectrum, exploring the underpinnings of blockchain's socio-economic repercussions and the metaverse's potential. The innovative aspect of our work lies in its amalgamation of technological, financial, and social dynamics within the realm of cryptocurrencies and the Metaverse. Conversely, the compared article offers a deep technical dive into the optimisation of clustering algorithms for targeted financial outcomes. In essence, while both pieces contribute valuably to the discourse on financial systems and risk, they illuminate different facets— one the technical intricacies and the other a comprehensive socio-economic and technological panorama.

## 3. Academic literature review.

Decentralised finance is at present isolated to Blockchain projects. Still, most of the existing financial instruments can be transferred into an alternative economic infrastructure with the help of Blockchain Oracles [34]. Decentralised finance transactions can be cleaner than traditional finance and support 'local climate adaptation planning and implementation' [35]. This was a natural progression, given that digital finance and renewable energy consumption have already been studied for economic growth and technological progress, with evidence from China [36]. Blockchain and digital finance have also been the subject of other studies [38]. Especially the 'risk of blockchain technology in Internet finance supported by wireless network' [37].

It has been almost half a decade since the Blockchain was suggested as the new 'secure, decentralised, trusted cyber infrastructure solution for future energy systems' [38]. One of the solutions proposed in 2019 was the new concept for improving industry with private capital in China, with renewables finance and investment [39]. However, these concepts are not immune to the earlier problems related to 'the governance of local infrastructure funding and financing' [40], especially the role of mitigating the effect of the recession. As the authors stated in 2015, 'austerity and the fiscal consolidation of public finances have reinforced government efforts to reduce expenditure and debt, and secure private sector engagement and resources [40].

Blockchain values are also studied in decentralised electricity access developed with private investment as a sustainable development finance business model [41]. There are various platforms for decentralised autonomous organisations [42], even in just one of the existing blockchains, such as Ethereum. Many new Blockchains, such as Algorand, Cardano, Solana, and Avalanche, are emerging daily. The most interesting are the new Blockchains with cross-chain operability, such as Cosmo, Polkadot, and Chainlink. There are also numerous 'automated market makers and decentralised exchanges' [43], some even for cryptocurrency trading [44]. The question is, do the new regulations 'provide legal certainty' [45]?

The proper form of decentralised finance will always be based on private investment. Still, solid regulations and guidance also support centralisation in financial instruments. Before transferring traditional instruments into a new system, we must analyse 'the return–volume relationship in decentralised finance' [46]. Although decentralised banking is not an entirely accepted model yet, even in the present adoption stage, digital banks 'can learn from decentralised finance' [47]. The European Union has recently advanced into new Blockchain regulations based on the 'Regulation on Market in Crypto-Assets' and 'Decentralised Finance - MiCA' [48]. The United States has also produced new regulations [16], [17], [49], and the United Kingdom is slowly catching up [20]. However, the countries that can benefit the most from new financial instruments are not the EU, the US, and the UK.

Developing African countries have been very flexible in adopting Blockchain projects, as recorded in the recent study on 'Decentralised Finance and Cryptocurrency Activity in Africa' [50]. The values from Blockchain projects and new Metaverses do not have to be purely financial. For example, Blockchain projects have been 'empowering school-based management through decentralised financial control' since 2008 [51]. Despite these best efforts, even in 2023, we are still working on the 'conceptualisation and outlook' of 'decentralised finance platform ecosystems' [52]. This is predominately because Blockchain projects, and decentralised finance, have been advancing and developing in various areas. One example is creating an asset-backed decentralised finance instrument for food supply chains in the livestock export industry [53].

Another example is the 'decentralisation on health-related equity' for the 'decentralised governance of health care' [54]. The most critical example is from the Monetary and Economic Department, which discusses the 'potential benefits and challenges of the new system and presents a comparison to the traditional system of financial intermediation' [55]. This signals that decentralised finance is becoming a mainstream topic.

# 4. Case Study Review: Case studies of existing Blockchain solutions.

## 4.1. Blockchain 3.0 and Web 3.0

Blockchain technology, initially introduced through Bitcoin, combines cryptography with distributed computing, both of which have existed for several decades. Blockchain 1.0 represents a distributed secure database where a network of computers collaborates and shares data. In this architecture, individual computers, acting as network nodes, validate transactions and propagate them to other network nodes, creating a blockchain. Blockchain relies on a distributed consensus algorithm, requiring agreement among different nodes before any alteration can be made. The interdependence of blocks is achieved through hash values, making it virtually impossible to delete a block without affecting the entire chain.

The evolution of blockchain technology has led to Blockchain 2.0, exemplified by Ethereum, which introduces the capability to execute any computer code on the system, essentially creating a distributed virtual computer. This advancement opens the doors to various applications, envisioning a decentralised Turing-complete virtual machine. Blockchain 2.0 enables the creation of decentralised ledgers for asset registries, encompassing physical and intangible assets, such as property, currencies, patents, votes, identity, and healthcare data. It replaces the need for multiple private databases with a shared, trusted database accessible by all relevant parties. The immutability of data stored on a blockchain enhances its credibility, as it is incredibly challenging to alter or corrupt.

Despite the potential of existing blockchain solutions, they are often considered inefficient and face scalability issues, leading to the emergence of the third generation, Blockchain 3.0. Prominent examples of this generation include IOTA and Dfinity. Blockchain 3.0 introduces "The Distributed Cloud," a pivotal infrastructure supporting the development of the next generation of the internet, also known as Web 3.0 or the decentralised web. A broader ecosystem of technologies is required, including advanced web data analytics and the Internet of Things (IoT). This integration of technologies enables the storage and analysis of vast amounts of sensitive data, unlocking new value and insights through cross-correlations from diverse IoT data sources integrated into the blockchain.

The actual value of IoT data lies not in making individual devices or systems bright, but in enabling seamless processes across domains, organisations, and procedures. This necessitates open networks capable of communicating and coordinating components on demand. Blockchain 3.0 and Web 3.0 mark a paradigm shift from ownership to "Servitization," where assets are used as services rather than owned outright. This shift requires the establishment of frictionless markets and automated exchanges, with the blockchain acting as the trust machine, ensuring secure and transparent transactions.

In summary, the case study review demonstrates the evolution of blockchain technology from its initial implementation in Bitcoin (Blockchain 1.0) to Ethereum and other platforms (Blockchain 2.0) and the ongoing development of Blockchain 3.0 solutions. These advancements present new opportunities for decentralisation, enhanced data analytics, and the integration of IoT, culminating in Web 3.0. By embracing these technologies, we can transition towards a future where assets are utilised as services, supported by seamless processes and trust-enabled exchanges facilitated by the blockchain.

## 4.2. IoT-based Blockchain solutions.

The Internet-of-Things (IoT) is already used in Blockchain 3.0 as an open-source distributed ledger (e.g., IOTA, NEO, EOS) and has presented many unique alternatives for storing transactions with a potential for higher scalability (by using Tangle) over Blockchain 1.0 based distributed ledgers (such as Bitcoin). However, the interest in some of the early crypto projects seems to be dying down. In Figure 5 we can see the main research interest from the pre-2018 bull run.

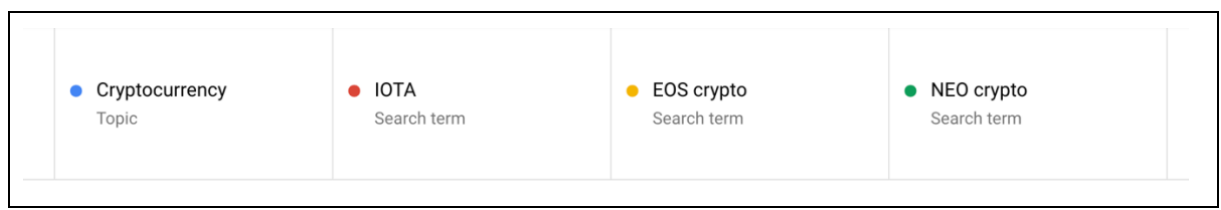

# University of Oxford

**Petar Radanliev, BA Hons., MSc., Ph.D.**
POSTDOCTORAL RESEARCH ASSOCIATE

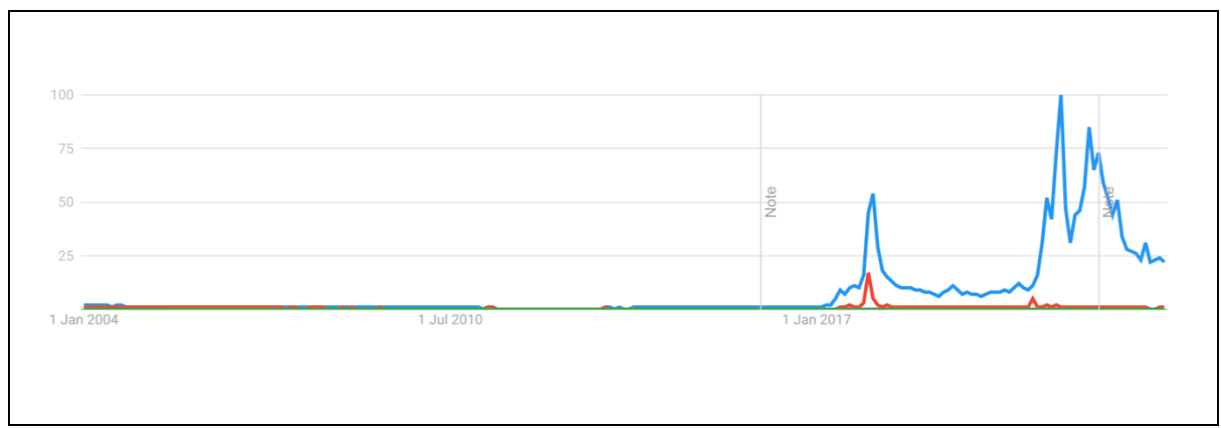

*Figure 5: Trends in interest for early (pre-2018) crypto projects vs interest in cryptocurrencies in general.*

One solution that seems to be under consideration is to rename existing projects (e.g., NEO is renamed to N3 – with a promise of a better Blockchain)[5]. IOTA is keeping its name but evolving into a new blockchain called Crysalis[6]. Although the idea of IoT-based crypto is undeniably valuable, further research is required to determine how IoT technologies would resolve the main problem of Blockchain 1.0, namely the (i) speed of transactions, and the (ii) security risk from quantum computers [56]–[62]. For the first part of the objective, the IOTA project can process around 500-800 transactions per second. Many faster blockchains like Solana can process 50,000 transactions per second. However, Solana and other Blockchain 2.0 are run by validators or entirely by volunteers. In other words, random people making impossible to guarantee that the network is altogether reliable. IOTA and other Blockchain 3.0 technologies are run by IoT systems, which can be considered reliable if the devices are secure. Security and reliability are necessary for technological adoption in critical infrastructure, and NHS supply chains can benefit from improved security that comes with Blockchain technologies. Still, scalability is a significant concern for the adoption of Blockchain 1.0. In other words, at present, Blockchain 1.0 can process up to seven cryptographic hash function computations per second and has a confirmation time of 15 minutes.

In comparison, Visa processes around 1,700 transactions per second on average (based on a calculation derived from the official claim of over 150 million daily transactions). The potential for Blockchain adoption is there but is currently bottlenecked by scalability. Another obstacle to the adoption of blockchain is the energy demand and the cost (fees).

Most Blockchain 1.0 (Bitcoin) and 2.0 (Ethereum) come with heavy fees, while Blockchain 3.0 (e.g., IOTA) can have zero fees. Zero fees are possible because IoT devices can communicate autonomously and send verified messages, enabling full automation of decentralised crypto, and future versions can make nano payments possible. The main problem with IoT Blockchains is the security of IoT devices. Until these devices are secure, it is difficult to see mass adoption of IoT-based crypto. The IOTA project has been in the cryptocurrency markets for a long time (in crypto terms) and still hasn't picked up in price or mass adoption.

---

[5] https://neo.org/

[6] https://chrysalis.iota.org/status

## 4.3. IoT for healthcare.

The increased adoption of IoT in health services seems inevitable [63]–[68] because IoT offers increased efficiency at a reduced amount of time spent, which appears to be what the NHS needs, but the value of the low cost of IoT comes at risk [69]–[71]. To reduce costs and provide the required support, the NHS must automate some essential data collection and monitoring, freeing skilled staff to focus on patient safety and patient service. This IoT automation in national critical infrastructure must come with increased (or at least matched) security to the traditional legacy IT systems [72]–[74], and there are various solutions for improved cybersecurity [75]–[81].

One of the leading crypto projects that could benefit NHS is the 'VeChain' project, which resolves many of the existing supply chain issues [82]–[123], especially the software supply chain, which has many areas of current work, including the Software Bill of Materials (SBOM) [124]–[128], the Vulnerability Exploitability Exchange (VEX) [129]–[131], and many other research efforts. The 'VeChain' project addresses many of these long-standing areas of concern and is already operational. Still, it would be interesting to have even more automated, machine-controlled supply chains in the future. For that, we have a very different set of Blockchain solutions – based on IoT systems [97], [132]–[138].

Using existing IoT Blockchain solutions in the NHS-specific requirements, multiple healthcare functions can be performed simultaneously. The main concern is the lack of interoperability. Designing specific solitons individually and rolling out such solutions piecemeal could result in a lack of interoperability between one NHS trust and another, or even one hospital and another in the same trust. The expected outcome of such blockchain solutions is an interoperable of-the-shelf solution explicitly tested for compatibility with existing NHS systems.

In addition, by using established IoT Blockchains solutions, the expected outcomes are resolutions to some challenging philosophical questions around the use of IoT in NHS, such as issues around data ownership. Since IoT Blockchains operate as decentralised systems, with cryptographic hashing of data in blocks, this will resolve whether the data belongs to patients, the NHS, and the technology providers. By storing personal data in Blockchain Cloud solutions, such as the NuNet Cloud, a centralised authority won't own the data, and sharing of the data, or alternative uses and analysis, could only be achieved by a consensus of the Blockchain validator nodes. Depending on what Blockchain is used, this could be many validator nodes.

## 4.4. Decentralised finance – DeFi.

Decentralised finance (DeFi) is built upon automated protocols that offer financial services. DeFi solutions bring numerous advantages compared to traditional finance, including faster transaction speeds, improved transparency, interoperability, and immutability. However, these strengths also give rise to specific weaknesses. In other sections of this review, we explore the challenges of ensuring compliance with anti-money laundering processes when conducting cross-border transactions at high speeds. The rapid pace of DeFi operations often comes at the expense of non-compliance. Although this issue is expected to be resolved in the future, it poses a significant hurdle for widespread adoption, particularly on a national level, as observed in the case of El Salvador.

In 2023, several significant DeFi protocols have emerged, such as Aave, Compound, Curve Finance, Synthetix, PancakeSwap, Kyber Network, Uniswap, and others. While new protocols are gaining momentum, Uniswap remains the most popular decentralised exchange. The collapse of centralised exchanges, including FTX, has created new market potential for decentralised exchanges. Nevertheless, questions persist regarding the level of decentralisation achieved by some of these platforms.

As DeFi continues to evolve, striking a balance between speed, compliance, and decentralisation becomes crucial for sustainable growth. Overcoming the challenges associated with regulatory compliance will facilitate the broader adoption and integration of DeFi into existing financial systems. Addressing concerns about the degree of decentralisation within decentralised exchanges will foster greater trust and confidence among users. By navigating these complexities, the DeFi ecosystem can unlock its full potential and reshape the accessibility of financial services, empowering individuals worldwide with greater financial inclusivity and autonomy.

## 4.5. Centralised exchanges.

Centralised exchanges faced significant challenges in 2022, and without comprehensive regulatory oversight, they appear poised to repeat past mistakes. While surviving centralised exchanges like Binance, Coinbase, Kraken, and KuCoin, among others, claim to adhere to compliance standards, the absence of effective oversight is evident, with limited exceptions in the United States, Australia, and New Zealand. In Australia and New Zealand, certain major exchanges are subject to regulatory measures that encompass ATO (Australian Taxation Office) and AML (Anti-Money Laundering) compliance, as well as KYC (Know Your Customer) protocols. Coinbase in the United States operates under a similar regulatory framework. However, the effectiveness of such compliance measures warrants further examination in a separate article. In this piece, we focus on the level of compliance necessary for crypto assets to be fully compliant.

The issue of partial compliance can be perplexing, as classifying inherently risky assets as compliant creates opportunities for fund managers to include these assets within seemingly secure financial products, such as pension funds. This scenario raises concerns about a potential future parallel to the subprime mortgage crisis 2008. Thus, it is crucial to thoroughly assess and understand the actual compliance status of crypto assets to avoid misleading categorisations that may inadvertently contribute to the creation of unstable financial products.

Implementing robust regulatory measures and ensuring effective oversight are imperative steps towards fostering transparency and safeguarding investors' interests within the cryptocurrency landscape. Striking a balance between innovation and regulation will be critical in building a sustainable and resilient financial ecosystem that mitigates risks and instils confidence among market participants.

### 4.6. Layer 1s and 2s, Non-Fungible Tokens (NFTs), and the Metaverses.

Blockchains such as Bitcoin and Ethereum are called Layer 1 in decentralised terminology. Alongside these Blockchains, there are Layer 2 protocols that can be utilised in conjunction, enhancing the capabilities of the underlying Blockchain. Examples of Layer 2 protocols include Arbitrum and Optimism. Within the Metaverse landscape, many platforms operate as Layer 2 protocols built upon existing Blockchains. For instance, Decentraland (MANA), ApeCoin (APE), Axie Infinity (AXS), the Sandbox (SAND), Enjin Coin (ENJ), Gala (GALA), Render (RNDR), and Metahero (HERO) are all powered by the Ethereum (ETH) blockchain. Theta (THETA) initially started as an ERC-20 token but subsequently transitioned to its native THETA token. Theta utilises two tokens: THETA for governance and TFUEL for utility. Stacks (STX) is one of the few layer-1 blockchains in the Metaverse ecosystem.

Non-Fungible Tokens (NFTs) represent unique pieces of art, digital content, or media within the Metaverse. These tokens enable trading and serve as a means of storing value. However, it is essential to note that not all NFTs have proven to possess the anticipated value that some investors had hoped for.

To summarise the information presented:

| Blockchain/Protocol | Layer Type | Notable Tokens |
|---|---|---|
| Bitcoin | Layer 1 | - |
| Ethereum | Layer 1 | MANA, APE, AXS, SAND, ENJ, GALA, RNDR, HERO |
| Arbitrum | Layer 2 | - |
| Optimism | Layer 2 | - |
| Theta | Layer 1 | THETA (governance), TFUEL (utility) |
| Stacks | Layer 1 | STX |

The Metaverse ecosystem encompasses a diverse range of blockchains, protocols, and tokens, providing unique opportunities for participation and engagement within virtual environments.

### 4.7. Crypto Bridges and Oracles.

Blockchain bridges enable the movement of assets from one Layer 1 to another Layer 1 or from Layer 1 to Layer 2 and the reverse. Some of the most famous bridges in 2023 include Hop Exchange, Orbiter, Rango Exchange, cBridge, xPollinate, AllBridge, and many others. Many of the major hacks that resulted in a significant loss of Crypto in 2022 were based on cyber-attacks on Crypto bridges. While Bridges resolve interoperability issues across chains, Oracles (such as Chainlink) enable cross-chain communication and intelligent contracts to execute on different Blockchains.

### 4.8. Crypto Wallets.

Cryptocurrency digital wallets can be classified into three main categories. The first category comprises safe storage tricky wallets, where cryptocurrencies are securely stored on a personal device. Examples of such wallets include Ledger and Trezor. The second category consists of hot wallets, such as Metamask and TrustWallet, software-based wallets typically accessed through internet-connected devices. Lastly, there are exchange wallets, such as Coinbase Wallet, provided by Centralised exchanges.

Considering the recent collapses of Centralised exchanges, it becomes challenging to understand why individuals still choose to store their cryptocurrencies in exchange wallets. Users may prefer the convenience of having someone else manage the day-to-day operations of their savings and finances. However, this reliance on Centralised exchanges highlights the importance of implementing robust regulations to ensure the security and integrity of these platforms.

To summarise the information presented:

| Wallet Type | Examples |
| --- | --- |
| Safe Storage | Ledger, Trezor |
| Hot Wallet | Metamask, TrustWallet |
| Exchange Wallet | Coinbase Wallet |

Classifying cryptocurrency wallets into these categories provides users with various options to suit their preferences for security, accessibility, and management of their digital assets. It is crucial to weigh the advantages and risks associated with each type of wallet to make informed decisions regarding the storage and protection of cryptocurrencies.

# 5. Lessons to be learned from the past errors: the two Cases of FTX and Terra Luna

FTX was a centralised cryptocurrency exchange, providing crypto derivatives and leverage trading services. Still, the primary use for centralised exchanges is to enable customers to buy and exchange different cryptocurrencies. The main problems resulting from the collapse of FTX also apply to all other centralised cryptocurrency exchanges currently in operation. Cryptocurrency exchanges are not regulated, which leads to individuals taking risks without the approval of the asset owners, which is an oversimplified description of why centralised cryptocurrency exchanges should not be allowed to operate without regulations. Individual savers are not always keen on keeping their assets on a USB drive or writing the private keys on a piece of paper, which, if lost, would result in the loss of their savings. Hence, many small crypto savers use centralised cryptocurrency exchanges (such as FTX) to store their crypto savings and earn interest – similarly to the traditional banking system. Despite all the warning signs, individual savers are still locking their crypto savings in unregulated centralised exchanges.

This presents three options to the UK government and all other governments worldwide.

1. The first is to create standards and regulations for cryptocurrencies because as of today (08 Jan 2023), we have 22,228 different cryptos (i.e., crypto projects) and 534 crypto exchanges, with a market cap of $824,468,428,103 and 24h trade volume of $16,374,071,351[139]. None of the trades are regulated in the UK, nor most other countries.

2. The second is to ban all use of cryptocurrencies, including ownership and trading, but this is unlikely to be effective because most crypto projects are run from outside of the UK, and some (e.g., Bitcoin) are decentralised. Hence, even if a global task force could be created to track and trace cryptocurrency projects and exchanges, it would be ineffective against decentralised crypto and will only push trade and ownership into the dark economy. In addition, it is unlikely that the legal mechanisms can cope with persecuting all cryptocurrency projects and exchanges because, as we can see from the case of the XRP legal proceedings, just one point can take years to resolve. The US Government has proven that it can effectively ban crypto projects. In August 2022, the U.S. Treasury sanctioned the virtual currency mixer Tornado Cash [140]. The Tornado Cash DAO was shut down, and its lead developer Aleksey Pertsev was arrested, but what this translates to is that the mixer's code itself is banned for use, and it does not mean that the code has been disabled and cannot be used. It means that the Tornado Cash U.S. crypto customers are not allowed to use the mixer, at least not without permission from the U.S. Treasury. The mixer is blacklisted in the US because of its use in money laundering. However, the Tornado Cash app will continue to operate on the Ethereum blockchain exists. The critical point is that it is impossible to shut down such technology without shutting down the entire Blockchain. Since some Blockchains are decentralised, this will prove difficult, and even, if possible, many new Blockchains are constantly emerging. Hence, sanctioning and banning are unlikely to be valid for completely closing all operations.

3. The third option is to create fully centralised Government run Blockchains, upon which open crypto projects and exchanges can be built. In this scenario, Governments could control the type of projects and impose regulations and standards upon the developers and the user community. In such fully centralised Blockchains, the government could allow the development of centralised and decentralised crypto exchanges and fund or encourage the development of CBDCs (Central Bank Digital Currencies) and regulated Stablecoins (cryptocurrency with a pegged value to another currency, commodity, or financial instrument). By enabling the development of a fully regulated Stablecoin, the UK Government would prevent one of the main risks for individual crypto savers: the collapse of another Stablecoin, which happened to the UST Algorithmic Stablecoin in 2022. Many of the current stablecoins are highly speculative, and at present most stablecoins are not audited or regulated – at least not in any meaningful way. Although Tether (USDT) has announced that it is preparing to be audited by a large accounting firm to prove the transparency of Tether, at present, USDT market reserves are not audited. As of today, Tether's market cap is $66,268,895,618. Around $11,106,992,770 of the cryptocurrency stablecoins traded in the last 24 hours alone. Tether (USDT) is just one of many stablecoins on the many current crypto exchanges. In the top 10 cryptocurrencies by market cap, apart from USDT, we also have the USDC (market cap of $43,922,152,193) and BUSD (market cap of $16,377,185,225).

4. In contrast, in 11th place, we have DAI (market cap of $5,790,436,026). In the 41st place, we have USDP (market cap of $876,254,775). On the 43rd place is TUSD (market cap of $846,271,617), in the 52nd place is USDD (market cap of $707,743,989) and so on – data from the 8th of January 2023 [139]. From the above-listed stablecoins, USDC has reserves regularly attested but not audited. None of the stablecoins are audited. This creates a systemic risk for all cryptocurrencies, and regulating the stablecoins will not only prevent future loss of savings for individual users and savers (hodlers), but it would increase the confidence in the market. Combined with a regulated crypto exchange, it would provide security and quick exit for investors during times of volatility. In the final comment on CBDCs, we must point out that the view emerging from this article is not sympathetic to the values of society and economy from CBDCs. Although CBDCs would resolve many issues related to fluctuations in the price of all cryptocurrencies, the stablecoin solution could be a preferred version of a Blockchain-based currency, specifically, decentralised stablecoins. However, the collapse of UST – LUNA has exposed vulnerabilities in some of the decentralised algorithmic stablecoins. We need new solutions to address some of the vulnerabilities disclosed in 2022.

5. The main lesson we must learn from FTX is that without taking regulatory action, corporate malfunction and malfeasance cases will continue to dominate the cryptocurrency ecosystems. Even if governments worldwide embrace the concept of complete monetary decentralisation (which seems highly unlikely), some crypto market elements still need to be regulated to ensure that self-governance is not replaced again with malfeasance. The collapse of FTX (which was considered one of the safest exchanges because of the public display of approval from various high-profile politicians), has proven that corporate malfeasance exists in cryptocurrencies on a much greater level than we are aware. To put this into perspective, if users start withdrawing large volumes from any of the above-listed stablecoins, it seems questionable if they will survive. That is not to say that the concept of stablecoins should be abandoned or that the currency should be pegged to gold and not to the USD. Stablecoins provide crucial services in the crypto markets, and USD is the most traded currency. The concept seems sound, but the regulations, standards and accountancy audits are missing.

These three scenarios could be seen as opportunities for the UK Government to intervene and take advantage of the situation to establish the UK as the leading country in the world, that is providing a highly demanded service (which is also highly profitable), but also highly regulated, standardised, and audited according to international standards. Would this expose the UK economy to unnecessary risk? That depends on how this process is undertaken.

Suppose the UK develops a new Blockchain that provides the services that companies use to build crypto projects. In that case, the risk to the UK is minimal, even if the crypto markets collapse. The value of Bitcoin dropped to its all-time lows from year 2009. The Blockchain would still charge transaction fees until the market cap goes down to zero, at which point, there won't be any transactions, and there won't be any cost, because there won't be any need for maintaining the new Blockchain. Similar arguments can be made about the development of a UK crypto exchange. Regardless of the level of centralisation, the code for Uniswap and many other exchanges are open and can be copied to create a new exchange without building a new code. That is precisely what SushiSwap did, gaining a significant market cap and trading volume almost instantly.

## 6. Survey of Crypto use cases.

The use case for crypto projects depends mainly on the specifics of the project features and characteristics. Some of the most popular use cases come with questionable motives. For example, the idea that bitcoin can be considered as digital gold, or as a store of value, and that bitcoin can be used to preserve wealth and hedge against inflation. This use case is very debatable. There are many use cases for crypto, and below I list some realistic use cases, but the idea that a digital asset with no other purpose or a use case can replace gold, is not very convincing. It seems more likely that Bitcoin will need to be wrapped and transferred to a different chain, where the cost of transactions is much lower, and be used as a payment system, similar to SWIFT.

That could be a real-world use case, and we already have the Algorand Blockchain, which is capable of handling wrapped Bitcoins, and the cost of transactions is very low, while security is high. Algorand could even enhance the security of Bitcoin. There will be many other Blockchains that can do the same function. Hence, the bitcoin community needs to start innovating because back in 2009, Satoshi presented the most innovative and secure technology, but surely, he didn't expect this to remain the same forever.

Some of the real-world use cases include:

1. Borderless payments without any centralised entities acting as an intermediary,
2. Decentralised finance for lending and borrowing, accessible to anyone and everyone that has an internet connection and knows how to use the specific blockchain–crypto project,
3. Security and privacy of products (and services) as they move through different supply chains,
4. Authenticity verification for products and services,
5. Ensuring payments are processed somewhat in supply chains- with the use of smart contracts,
6. Buying and owning digital assets, such as gaming and collectables, can be held as non-fungible tokens (NFTs).

7. Crypto transactions can be designed to help with privacy and anonymity, creating added value for users that prefer to keep their finances private.
8. Crypto can be used for crowdfunding, where new funds can be raised by issuing initial coin offerings (ICO) or token generation events (TGEs).
9. Creatives can use crypto to monetise their work. For example, digital content creators can accept payments for premium content, opening a safer and cheaper environment for various artists, from dancers, tutors, and painters to fitness instructors, digital consultants and education providers.
10. Charitable donations, crypto can be used as a fast and secure method for transferring wealth to people in need.

New use cases will continue to emerge with the increased adoption of decentralised blockchain technology. One way to compare the current state of crypto is to think of web one vs web two and the emergence of web three. Web one was just a collection of data and information made available for free online.

## 6.1. Evolution and Uncertainties of the Internet: From Web1 to Web3

The internet has undergone significant transformations over the years, from the emergence of Web1 as the "information highway" to the interactive capabilities of Web2. However, Web2 has raised concerns regarding privacy exploitation, paving the way for the anticipated arrival of Web3, which is expected to be built on blockchain technology. The blockchain's primary use case has already found its place in adopting blockchains within Web3. This section explores the evolution of the Internet and the uncertainties surrounding the dominance of specific platforms and cryptocurrencies.

The transition from Web1 to Web2 can be likened to the shift from AOL (the initial popular web browser) to Hotmail. Similarly, Ethereum has emerged as the dominant force in the crypto world, particularly with the introduction of layer two projects like Optimism and Arbitrum. Ethereum's upgraded blockchain, transitioning from proof of work to proof of stake, became a prominent player in 2023. However, whether Ethereum will sustain its popularity in the coming years or if another cryptocurrency will surpass it. This draws parallels to Hotmail's decline despite experimenting with features similar to its competitors.

The similarity between the current version of WhatsApp and the older MSN chat is quite puzzling. While WhatsApp remains popular, MSN lost many users and was eventually abandoned. Although some argue that WhatsApp relies on mobile signals, it uses personal mobile numbers as usernames rather than a communication method. In contrast, MSN offered the advantage of creating new and random usernames, making it seemingly more secure. This example highlights the unpredictability of platform preferences among users.

As the emergence of Web3 unfolds, multiple platforms aim to provide similar services. Identifying the platform that will prevail in the long run is challenging. Ethereum may be the MSN in this scenario, but who will be the WhatsApp? It's not a single platform; projects like NEAR and SOL, among others, offer comparable functionalities to Ethereum. The long-term viability of crypto projects depends on their ability to deliver unique services and align with user preferences.

The question arises whether Bitcoin will experience a revival with the Lightning network. When transmitted via the Lightning network, there is no real distinction between layer one and layer two solutions. Additionally, Bitcoin has been striving to increase the use of renewable energy and improve speed and user-friendliness for everyday transactions. Perhaps, there is potential for new blockchain innovations to emerge from Bitcoin.

In summary, the Internet has evolved from Web1 to Web2, and the anticipated arrival of Web3 brings new uncertainties regarding platform dominance and the emergence of innovative cryptocurrencies. The future landscape of the internet and blockchain technologies remains unpredictable, as users' preferences and advancements in various projects play pivotal roles in shaping the future of the digital landscape.

### 6.2. The Buterin's trilemma.

This review article would not be complete without discussing the Buterin's trilemma and how that fundamentally captures the trade-offs between security, decentralisation, speed, and the attendant risks. Some blockchains (e.g., Sol) go for speed but are more centralised; others are less centralised but often more secure. One study suggested a solution called *'The Blockchain Quadrilemma'* but also recognised that for basic Blockchain operations *', Algorand can often be the right choice'*, but Ethereum is recommended for *'more sophisticated computations'* [141]. Another research study recommended *'a dichotomy of algorithms between leader-based and voting-based consensus algorithms'* based on *'tradeoffs … for a given distributed system'* [142] It is worth mentioning that some experts recommend an *'incentive-based role in the governance of DeFi as opposed to an enforcement-oriented role'* because we need to build new tools that enable new *'policy options in a transnational environment hostile to formal state intervention'* [143].

## 7. Discussion.

Crypto is still subject to many risks for investors and users. One of the main risks for investors is that the value is highly volatile and fluctuates significantly in short periods. This makes crypto price and value extremely difficult to predict, leading to significant losses when the value drops significantly. Since crypto is also a speculative asset, investors cannot be confident that the value of their investment will ever recover or go to zero.

Another risk is that no one, a government, or any financial institution regulate crypto. This is the clearest indicator that there won't be any protection or oversight from fraud, mismanagement, or financial malfeasance. In other words, crypto investors need to be aware that they will have very little recourse if something goes wrong.

Apart from these risks, one commonly discussed risk is the safety and security of blockchain technology. The common topic in media is that the underlying blockchain technology is not as secure as it was thought. The rationale for this assumption is that in the past, we had some very high-profile hacks that resulted in the theft of large amounts of crypto from exchanges and wallets. Although the latest is correct, the first is not. Bitcoin and other secured blockchains have never been hacked, the hacks happen on crypto exchanges and digital wallets that do not apply appropriate security, and most of the theft has been on crypto bridges, pools, and other instruments unrelated to the blockchain technology itself.

To explain this, in other words, the Blockchain is as secure as it has been described in the first paper written by Satoshi, but the new projects are bypassing the security requirements, often because there are no cybersecurity standards. The hacks are increasing, but this does not mean that blockchain technology is not secure. It means that we need cybersecurity standards for projects that use blockchain technology.

## 7.1. The good.

Traditional finance has also been slowly developing new, faster and more secure solutions. The SWIFT network is very slow compared to some of the crypto solutions, and the idea of tokenised USD does seem appealing to many users, specifically traders. Cryptocurrencies can also solve many of the banking problems in developing countries. The ability to make payments and transfer tokens pegged with the value of USD could provide solutions to much of the developed world still lacking essential banking services. Since existing payment methods like Visa or Mastercard charge service fees, it is reasonable to expect that some small fees would be acceptable to users. However, there is no evidence of crypto being adopted as a payment method in developing countries. Although some countries like El Salvador have adopted Bitcoin, its use for everyday payment has not been adopted. Another point to make here is related to the value of Bitcoin (BTC) and private money and wallets. Although decentralised cryptocurrencies don't hedge against short-term inflation, Bitcoin has massively outperformed gold over a decade despite monetary easing/printing.

## 7.2. The bad.

Bitcoin emerged from the financial chaos in 2008, and it was presented as a solution to the centralised banking system and the high-risk practices of a few greedy financial firms that only care about profits. The cryptocurrencies only replaced one centralised set of intermediaries that are strictly regulated, with another set of centralised intermediaries that are not regulated. Despite the fall of FTX, crypto exchanges do not want to be controlled, and they are challenging to regulate, because they can be run from anywhere in the world.

One of the reasons for this is that consumers prefer the safety and service of third parties looking after their money and rarely prefer keeping their savings in a cold wallet that can be lost, damaged, or hacked, along with all the money saved. The shift from decentralised money to centralised exchanges just shows that people prefer the convenience of a third party looking after their money.

The long-term economic utility is also questionable. Why would one coin created out of code be valued in thousands of pounds and another coin created out of code be worth zero. It is often compared with traditional securities such as stock shares, but stocks generate cash flow, and we can discount the cash flow to the present time to come up with a valuation. Another example is that a fiat currency is valued relative to other fiat currencies based on GDP, inflation, interest rates, and other data from different countries. None of these valuations apply to cryptocurrencies.

This means that cryptocurrencies cannot be valued because they do not have any trade fundamentals. Instead, cryptocurrencies trade purely on sentiment, and most of the price spikes are created by influencers and social media. If this trend continues, we can argue that even decentralised cryptocurrencies are just decentralised Ponzi schemes and rely on a supply of new buyers to buy new tokens at higher prices. The supply of new buyers will run out eventually. Worth emphasising here is that from the five most famous Ponzi schemes of the 2020s, are allocated to cryptocurrencies (QuadrigaCX, Terra Luna, and FTX[7]).

What is really striking in this scenario is that the poorest usually learn last about the scheme and tend to lose the most. A clear example of this is the case of El Salvador, where they adopted Bitcoin as a legal tender, followed by credit rating agencies downgrading their sovereign credit rating. The IMF has started to cut off funding, nobody is using Bitcoin there to buy anything, and it turned out to be a disaster. Other developing countries have found effective fintech solutions that do not require crypto, such as the M-Pesa in Africa, which is based on SIM card payment from a mobile phone, or We Chat Pay, a QR Code payment. Both examples have proven success in adoption and financial inclusion, with billions of users and almost no banking infrastructure investment. If we compare this to Bitcoin, we are still not able to walk into a shop and buy things with Bitcoin or any of the cryptocurrencies that we have at present. It is questionable if we will ever be able to do that because of the cost of validating the transaction – it might simply not be viable to process so many transactions with the current Bitcoin mechanisms. It seems more likely that Bitcoin would need to be wrapped as a token on a different – less secure Blockchain just to be used as a currency for payment of everyday things, like coffee or beer.

For crypto to be seen as a long-term value, it must provide economic utility, and at present, crypto is not providing such utility. Another problem is that in technological terms, crypto is old technology, the first block was created in 2009, and we have seen a rapid technological change since then. Most of the mobile phones from 2009 are now considered old and almost obsolete. We are still waiting for some extraordinary use case for crypto, but since it hasn't materialised until now, the question is when it will, and would it ever happen?

### 7.3. The ugly.

Currently, the crypto market is not connected to traditional markets and is relatively small. If the crypto market relates to the traditional finance, the spill-over effect must be considered. The main concern is that if the crypto market is regulated in a way that it would get supercharged and it's allowed to create the connections between regulated finance and the crypto system, then crypto problems can become much bigger problems. This would mean that people that never invested in crypto are affected by price fluctuations and crypto market risks, in the same way that investors were affected by the mortgage-backed securities in 2008. Hence, the focus in regulations should be on minimising the connections between the crypto market and the regulated financial system.

---

[7] https://en.wikipedia.org/wiki/List_of_Ponzi_schemes

One significant risk from cryptocurrencies to our society is gambling, including leveraged and other forms of trading. In this article, we focused on the crypto ecosystem and the risk and values of the technology for society and the economy – from a perspective of improved wealth of a nation. But we also noticed that in the literature, we seem to have ignored that trading and gambling are addictions that humans have been subjected to since the early age of humanity, and the gambling market is booming in crypto. Wherever there is a demand, there will be a supply, and this needs to be regulated because gambling is a severe addiction.

One of the more concerning discoveries in this area is that Crypto gambling is *'distinct with regard to higher novelty seeking, higher gambling tendencies, and unique investment patterns'* [144] and this conclusion has been reached with the use of established and new Crypto specific methods, including 'Fear of Missing Out (FoMO) scale, Temperament and Character Inventory-Revised-Short (TCI-RS), Mood Disorder Questionnaire (MDQ), trait anxiety part of the State-Trait Anxiety Inventory (STAI-T), and the Korean version of the Canadian Problem Gambling Index (K-CPGI)'.

Cryptocurrency trading and gambling have also been associated with mental health problems, including depression and anxiety, with the main findings confirming a direct similarity between the demographic and personality characteristics of cryptocurrency traders and gamblers [145]. However, the study recognises that there could be 'differences between long-term investors and short-term traders of cryptocurrency', although, given the market uncertainty, some of the long-term investors in Terra Luna of FTX might disagree. Another recent study found that cryptocurrency trading results in 'rise to excessive or harmful behaviour including over-spending and compulsive checking', and although they identified many 'similarities between online sports betting and day trading', there are also some even more concerning factors, like 'the continuous 24-hour availability of trading, the global nature of the market, and the vital role of social media, social influence and non-balance sheet related events as determinants of price movements' [146].

These examples illustrate that we have reviewed some of the leading and most recent journal papers on this topic, but this topic needs much deeper research and understanding from a mental health perspective.

## 8. Conclusion.

In conclusion, this article has comprehensively examined the current state of cryptocurrencies and blockchains, aiming to enhance coherence, structure, and comprehensibility within the domain. We have shed light on the values and risks associated with these digital assets by presenting an up-to-date snapshot of the crypto landscape in 2023 and tracing its historical development from Satoshi's pioneering work. Moreover, we have endeavoured to clarify the distinctions between cryptocurrencies and blockchain technologies, addressing pertinent research questions regarding the innovative nature of blockchain, the significant risks posed by cryptocurrencies, and the potential societal and economic benefits of pursuing these technologies. Furthermore, we have explored the varying impacts on developed and developing countries and contemplated the longevity of different blockchain projects. While acknowledging the prevalence of fraudulent schemes, we have shifted our focus to the practical applications of blockchain projects, ultimately affirming the enduring presence of blockchain technologies. Our comprehensive discussion on the value derived from blockchain projects underscores their significance while re-evaluating key risks, including a personal reflection from the author on the potential risks.

The question that emerges from this review paper is, if we fast forward ten years, would all the transactions that we perform in our society be in fiat currencies, and the answer is that most probably they won't. With the emergence of Web3, assets, money and marketplaces will become interconnected, and some of the cryptocurrencies will form part of these new digital assets, but would that be Bitcoin, or some of the other 22,250 cryptos that are in circulation today, that is difficult to predict.

What also becomes clear is that regulation would eliminate many of the cases of corporate malfeasance. Regulations will most likely also remove many current use cases for crypto. Much of the recent hype around crypto is around the lack of regulations, and when regulations are applied, the promise of getting rich from crypto will certainly start to weaken. With regulations, crypto projects will have to start making checks (KYC) on who their customers are, and this argument for regulations killing the crypto is especially strong for cryptocurrencies that are created with no actual use case and based purely on the promise of making a great deal of money for early investors. Once regulations are created, these cryptos will be out of the picture. Most of the crypto projects are almost certainly not compliant with the derivatives or either security regulators.

Crypto has been operating in a very grey area, where different crypto project is considered as commodities, and because of that, crypto exchanges do not need to register with the federal government and be subject to regulation. This debate has been ongoing for far too long. The issue of whether crypto is a commodity or crypto is a security is not the main point of concern.

The main concern is not the naming but whether crypto is subject to regulation, and from this perspective, it makes sense to call all crypto assets securities, which will mean that all crypto is subjected to robust oversight. The issue is that, if that happens, most crypto projects won't be able to comply, which will hurt not only the crypto industry but also the crypto investors. Given that regulations are designed to protect investors, it is uncertain if such robust approach would serve the purpose it intended to, or would it lead to a significant loss for crypto investors. A more realistic approach would be to regulate crypto exchanges and ensure that exchanges are registered as investment dealers.

This seems realistic and reasonable because if a crypto investor engages in a contract with a crypto exchange, where the exchange would promise a very lucrative return or some exceptional benefits that are not very realistic, there are small crypto investors that might fail for such advertisements. This makes crypto exchanges investment deadlier, and they need to be regulated. Regulators need to engage in how these exchanges keep their assets, how they get the returns, and ensure that exchanges are not taking unnecessary risks that expose investors to risks they are unaware of or do not fully understand. The old saying in crypto is 'not your keys, not your crypto', and regulating the centralised exchanges, won't even come with any disagreement from the crypto community.

## 8.1. Final comments.

A final comment on investing in crypto, despite the remarkable returns some investors have benefited from, the crypto market has not matured yet, and the market – including the technology – is still evolving. There are risks, and investing in crypto comes with serious risks, and investors should approach crypto with caution. In addition, we still cannot use Bitcoin or any other Crypto to buy coffee in Costa or Starbucks, we cannot buy food in Tesco, and transactions come at a cost, while using fiat doesn't cost.

Until we can use crypto as a fiat currency in every aspect of the use cases, we won't be able to claim that retail and commerce adoption is increasing. Even if the price of crypto goes up (or down), the use case doesn't change much, which triggers concerns. On the other hand, the claim that Bitcoin is used for crime and money laundry has been contradicted by blockchain analysis company Chainalysis, which reported that only a very small percentage (0.15% total of $14bl) of known cryptocurrency transactions conducted in 2021 were involved in illicit activities [147].

# 9. Abbreviations

**Bitcoin** – the first decentralised blockchain.

**Terra Luna** – collapsed crypto project.

**FTX** – collapsed crypto exchange.

**Solana (SOL)** – crypto project that got affected by the FTX collapse.

**Ethereum, Cardano, Dogecoin, Litecoin, Algorand, NEAR** – crypto projects that remained popular with investors in the 2021 bull run.

**IOTA, NEO, EOS** – crypto projects that were popular in the previous bull runs, and are still present in the crypto market in 2023.

**NEFD** - new and emerging forms of data.

**CBDCs** - Central Bank Digital Currencies.

**Tornado Cash DAO** – crypto mixer that has been prohibited for use by the USA

**UST** Algorithmic Stablecoin – collapsed stablecoin.

**USDT, USDC, DAI, BUSD, USDP, USDD, TUSD** - Stablecoins still in existence in 2023 (as of 25th of January 2023).

**NHS** – National health Service.

**Uniswap / SushiSwap** – decentralised exchanges.

**NFTs** - non-fungible tokens.

# University of Oxford

**Petar Radanliev, BA Hons., MSc., Ph.D.**
POSTDOCTORAL RESEARCH ASSOCIATE

**ICO** - initial coin offerings.

**TGEs** - token generation events.

**ChatGPT** – AI based chat designed to replace the Google search engine.

**M-Pesa** - SIM card payment from a mobile phone.

**We Chat Pay** - a QR Code payment.

**KYC** – know your customer.

# University of Oxford

**Petar Radanliev, BA Hons., MSc., Ph.D.**
POSTDOCTORAL RESEARCH ASSOCIATE

# University of Oxford

**Petar Radanliev, BA Hons., MSc., Ph.D.**
POSTDOCTORAL RESEARCH ASSOCIATE